\documentclass[a4paper,11pt]{article}
\usepackage{paperstyle}

\usepackage{amsmath,amssymb,graphicx}
\usepackage{soul}
\usepackage[latin1]{inputenc}
\usepackage{dsfont}

\usepackage{subfigure}

\usepackage{array}

\newcommand{\be}{\begin{eqnarray}}
\newcommand{\en}{\end{eqnarray}}
\newcommand{\nn}{\nonumber\\}

\newcommand{\mpl}{m_{\rm{pl}}}
\newcommand{\As}{A_{\rm{s}}}
\newcommand{\ns}{n_{\rm{s}}}
\newcommand{\as}{\alpha_{\rm{s}}}
\newcommand{\ks}{\kappa_{\rm{s}}}

\DeclareMathOperator{\dphi}{d\phi}

\begin{document}

\begin{titlepage}

\vspace*{-15mm}
\vspace*{0.7cm}

\begin{center}

{\Large {\bf False vacuum energy dominated inflation \\[1mm] with large $r$ and the importance of $\ks$}}\\[8mm]

Stefan Antusch$^{\star\dagger}$\footnote{Email: \texttt{stefan.antusch@unibas.ch}},  
Francesco Cefal\`{a}$^{\star}$\footnote{Email: \texttt{f.cefala@unibas.ch}}, 
David Nolde$^{\star}$\footnote{Email: \texttt{david.nolde@unibas.ch}} and 
Stefano Orani$^{\star}$\footnote{Email: \texttt{stefano.orani@unibas.ch}}

\end{center}

\vspace*{0.20cm}

\centerline{$^{\star}$ \it
Department of Physics, University of Basel,}
\centerline{\it
Klingelbergstr.\ 82, CH-4056 Basel, Switzerland}

\vspace*{0.4cm}

\centerline{$^{\dagger}$ \it
Max-Planck-Institut f\"ur Physik (Werner-Heisenberg-Institut),}
\centerline{\it
F\"ohringer Ring 6, D-80805 M\"unchen, Germany}

\vspace*{1.2cm}

\begin{abstract}
\noindent
We investigate to which extent and under which circumstances false vacuum energy ($V_0$) dominated slow-roll inflation is compatible with a large tensor-to-scalar ratio $r = {\cal O}(0.1)$, as indicated by the recent BICEP2 measurement.  With $V_0$ we refer to a constant contribution to the inflaton potential, present before a phase transition takes place and absent in the true vacuum of the theory, like e.g.\ in hybrid inflation. Based on model-independent considerations, we derive an upper bound on the possible amount of $V_0$ domination and highlight the importance of higher-order runnings of the scalar spectral index (beyond $\as$) in order to realise scenarios of $V_0$ dominated inflation. We study the conditions for $V_0$ domination explicitly with an inflaton potential reconstruction around the inflaton field value 50 $e$-folds before the end of inflation, taking into account the present observational data. To this end, we provide the up-to-date parameter constraints within $\Lambda$CDM + $r$ + $\as$ + $\ks$ using the cosmological parameter estimation code Monte Python together with the Boltzmann code CLASS.

\end{abstract}
\end{titlepage}

\tableofcontents

\section{Introduction}

Recently, the BICEP$2$ collaboration reported on a measurement of the B-mode polarization of the CMB \cite{bicep2}, which may be interpreted as primordial gravity waves due to vacuum fluctuations of the metric during inflation, corresponding to a comparatively large tensor-to-scalar ratio $r \sim 0.2$. Although it is still too early to rule out alternative explanations such as primordial gravity waves from other sources \cite{BICEP:axion1,BICEP:axion2,BICEP:axion3}, topological defects \cite{BICEP:defects1,BICEP:defects2}, primordial magnetic fields \cite{BICEP:magnetic}, or large foregrounds \cite{Mortonson:2014bja,Flauger:2014qra}, it is worthwhile to study the implications such a large $r = {\cal O}(0.1)$ would have.

An immediate consequence would be that this large $r$ fixes the energy scale of single field slow-roll inflation to be 
\be
E_* = (V_*)^{1/4} = \left( \frac{3}{2}\pi^2 \As \, r \, \mpl^4\right)^{1/4} \simeq 9\times10^{-3}\mpl \simeq 2\times10^{16}\:\mbox{GeV} ,
\en 
where $\As\sim2.3\times10^{-9}$ is the amplitude of the primordial scalar perturbations. Interestingly, this energy scale is the same as the scale $M_\mathrm{GUT}$ where the couplings of the gauge interactions of the Standard Model (SM) of particle physics meet (when the minimal supersymmetric extension of the SM is considered), and where they could thus be unified in the context of a Grand Unified Theory (GUT). 

Both scales, $(V_*)^{1/4} $ and $M_\mathrm{GUT}$, may be the same just by coincidence, but there might also be a deeper physics reason behind it: If inflation was connected to a phase transition in particle physics, especially to the spontaneous breaking of the gauge symmetry of a GUT to the one of the SM, then the vacuum energy $V_0$ present before this symmetry breaking, to which we refer as the false vacuum energy $V_0$, might be identified with $M_\mathrm{GUT}$. This $V_0$ would contribute to the total vacuum energy $V_*$ at about 50 $e$-folds before the end of inflation. 

However, in addition to $V_0$, there is also a contribution to $V_*$ from the field-dependent part of the potential, which we name $\tilde V(\phi)$ with $\phi$ being the inflaton field. In this paper, we are interested in the situation in which the $V_0$ contribution to $V_*$ dominates over the one from $\tilde V(\phi)$, which would allow to explain the origin of the measured $V_*$ from a particle physics phase transition. However, this is not so easy to achieve: Large $r = {\cal O}(0.1)$ implies that the slope of the inflaton potential is comparatively large (at least at $\phi_*$, 50 $e$-folds before the end of inflation). This, together with the fact that (assuming slow roll inflation) the inflaton moves ${\cal O}(\mpl)$ during inflation, suggests that the vacuum energy contribution from $\tilde V(\phi)$ should also be sizeable. 

Models of inflation connected to particle physics phase transitions, such as hybrid inflation, tribrid inflation or models of plateau inflation have been studied in detail in the literature \cite{Lyth:1998xn,Mazumdar:2010sa,Martin:2013tda}. However, they are usually discussed with small inflaton field excursions $\Delta \phi$. In these scenarios $V_0$ generically dominates during inflation, but with the small $\Delta \phi$ they only feature a small tensor-to-scalar ratio $r \ll 0.1$. The question how large $r = {\cal O}(0.1)$ may be obtained in specific models of  hybrid inflation has been discussed in previous studies (see e.g.\ \cite{Carrillo-Gonzalez:2014tia,Kobayashi:2014rla,Choi:2014dva,smallField3,Pallis:2014xva}). 

In this paper, we investigate to which extent and under which circumstances $V_0$ dominated inflation is compatible with a large tensor-to-scalar ratio $r = {\cal O}(0.1)$ in a model-independent way. The paper is organised as follows: In section~\ref{sec:notations} we clarify our notation. 
Section~\ref{sec:analytic} discusses the relation between $V_0$ domination, relatively small inflaton field excursions $\Delta\phi$, and large runnings of the spectral index $\ns (k)$.
Then in section~\ref{sec:maxVacuumDomination} we derive an upper bound for the maximum possible $V_0$ domination in slow-roll inflation. 
In section~\ref{quartic} we study how the scale-dependence of the spectral index affects the amount of $V_0$ domination explicitly with a reconstruction of the potential around the inflaton field value at horizon exit. Finally, in section~\ref{sec:conclusions} we summarise and conclude.

\section{Notations and slow-roll formulas}
\label{sec:notations}

Throughout this paper, we use natural units $\hbar = c = \mpl = 1/\sqrt{8\pi G} = 1$, though we sometimes write the reduced Planck mass $\mpl$ explicitly to emphasise the mass dimension in some formulas.

The first four slow-roll parameters are defined as \cite{slowrollExpansion}
\begin{subequations}
\begin{align}
 \varepsilon \, &= \, \frac{1}{2}\left( \frac{ V' }{ V } \right)^2, \label{eq:defEpsilon}\\
 \eta \, &= \, \frac{V''}{V}, \label{eq:defEta}\\
 \xi^2 \, &= \, \frac{V'V'''}{V^2}, \label{eq:defXi}\\
 \sigma^3 \, &= \, \frac{(V')^2 V''''}{V^3}, \label{eq:defSigma}
\end{align}
\end{subequations}
where primes denote derivatives with respect to $\phi$. One can also define higher-order slow-roll parameters, but those will not be used explicitly in what follows.

The primordial spectrum is generated around $N_e \sim 50$ $e$-folds before the end of inflation; we indicate variables evaluated at that time by a subscript $*$. Analogously, variables evaluated at the end of inflation are denoted by a subscript $e$.

It is customary to write the spectrum of primordial curvature perturbations $\mathcal{P}_{\rm{s}}(k)$ in terms of an amplitude $A_s$ and a spectral index $\ns(k)$:
\begin{subequations}
\begin{align}
A_{\rm s} \, &= \, \mathcal{P}_{\rm{s}}(k_*),\\
\ns(k) \,&= \, 1 + \frac{d \ln \mathcal{P}_{\rm{s}}}{ d \ln k }, \label{eq:defNs}
\end{align}
\end{subequations}
with an arbitrary pivot scale $k_*$, which we choose as $k_* = 0.05$ Mpc$^{-1}$. The spectral index $\ns(k)$ is often expanded as a power series in $\ln(k)$, so that the spectrum $\mathcal{P}_{\rm{s}}(k)$ can be written as
\begin{align}
\ln \mathcal{P}_{\rm{s}} \, &= \, \ln A_{\rm s} + \left( \ns-1 \right) \left( \ln \frac{k}{k_*} \right)  \,+\, \frac{\as}{2} \left( \ln \frac{k}{k_*} \right)^2 \,+\, \frac{\ks}{6} \left( \ln \frac{k}{k_*} \right)^3 \,+\, ... ,
\end{align}
with the definitions
\begin{align}
 \ns \, \equiv \, \ns(k_*), \quad\quad \as \, \equiv \, \left. \frac{ d \ns }{d \ln k} \right|_{k_*}, \quad\quad \ks \, \equiv \, \left. \frac{ d^2 \ns }{(d \ln k)^2} \right|_{k_*}.
\end{align}
$\as$ is called the running of the spectral index and $\ks$ is called the running of the running. Note that throughout this paper, $\ns$ without an argument refers to the constant term $\ns = \ns(k_*)$, while $\ns(k)$ means the full function as defined in eq.~\eqref{eq:defNs}.

Analogously, one can define an amplitude $A_{\rm t}$ and a spectral index $n_{\rm t}$ for the tensor power spectrum $\mathcal{P}_{\rm{t}}(k)$. However, one usually uses the tensor-to-scalar ratio $r = A_{\rm t}/A_{\rm s}$ instead of the tensor amplitude.

In the slow-roll approximation, the primordial spectrum to leading order in the slow-roll parameters can be calculated as \cite{higherSlowroll,higherSlowroll2}\footnote{Eqs.~\eqref{eq:r}--\eqref{eq:kappas} assume Bunch-Davies initial conditions. For a discussion of non-Bunch-Davies initial conditions, see e.g.~\cite{BICEP:nonBD1,BICEP:nonBD2,BICEP:nonBD3,BICEP:nonBD4}.}
\begin{subequations}
\begin{align}
 r \, &= \, 16 \varepsilon_*, \label{eq:r}\\
 \ns \, &= \, 1 - 6\varepsilon_* + 2\eta_* + 2q_1 \xi^2_* + 2q_2 \sigma^3_* + ..., \label{eq:ns}\\
 \as \, &= \, -2\xi^2_* - 2q_1 \sigma^3_* + ..., \label{eq:alphas}\\
 \ks \, &= \, 2\sigma^3_* + ... \label{eq:kappas},
\end{align}
\end{subequations}
where $q_1 \simeq 1.063$, $q_2 \simeq 0.209$, and the dots denote slow-roll parameters involving higher derivatives of $V(\phi)$.

As we can always redefine our inflaton field as $\phi \rightarrow \pm(\phi - \phi_*)$, we can choose the sign $V'(\phi) < 0$ during inflation and $\phi_* = 0$. Together, these imply that we always have $\phi \geq 0$.

To leading order in the slow-roll parameters, one can then write the slow-roll equation of motion as\footnote{
Eq.~\eqref{eq:dNdphi} is valid for small slow-roll parameters $\varepsilon \ll 1$ (so that the potential energy dominates over the kinetic energy) and $\eta \ll 1$ (so that $\ddot{\phi}$ can be neglected during inflation), if the initial velocity  $\dot{\phi}_*$ is close to the inflationary attractor solution given by eq.~\eqref{eq:dNdphi}. We assume that these conditions are always satisfied during the first $N_\text{obs}\sim 8$ $e$-folds of inflation, so that eq.~\eqref{eq:dNdphi} is a good approximation to the full Friedmann equations.
}
\begin{align}
 \frac{dN}{d\phi} \, = \, -\frac{V(\phi)}{V'(\phi)} \, = \, \sqrt{\frac{1}{2\varepsilon}}. \label{eq:dNdphi}
\end{align}

We also need a precise definition of false vacuum energy domination during inflation, for which we write the inflaton potential as
\begin{align}
 V(\phi) \, = \, V_0 + \tilde{V}(\phi),
\end{align}
with $\min(\tilde{V}) = 0$, where $V_0$ is the false vacuum energy related to some GUT-scale phase transition which terminates inflation in the spirit of hybrid inflation. Throughout this paper, false vacuum energy domination refers to the condition that $\tilde{V} \ll V_0$ throughout inflation, or equivalently $V \simeq V_0 \simeq V_*$.

\begin{figure}[tbp]
  \centering
$\begin{array}{ccc}
\includegraphics[width=0.48\textwidth]{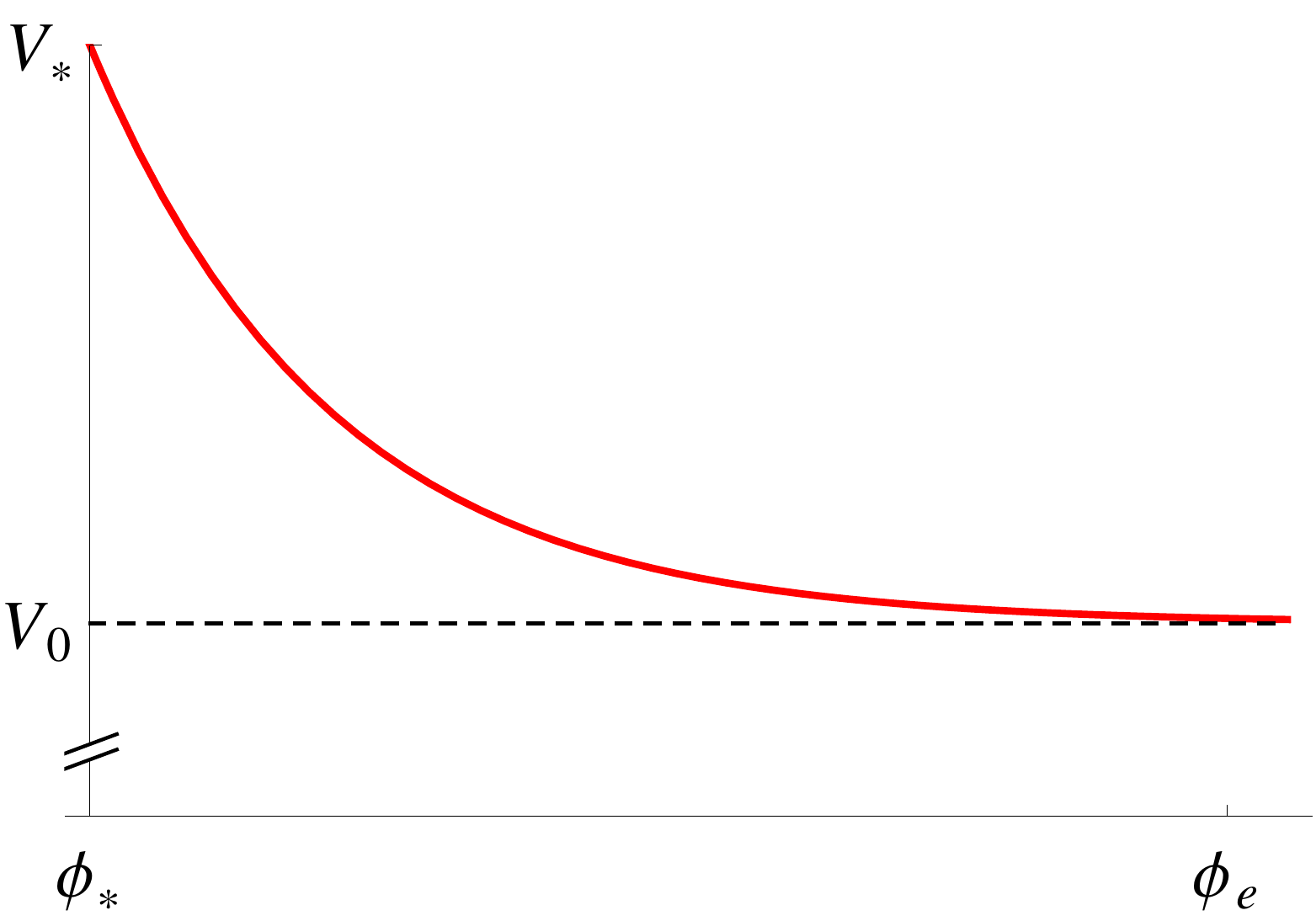} & \hfill &
\includegraphics[width=0.48\textwidth]{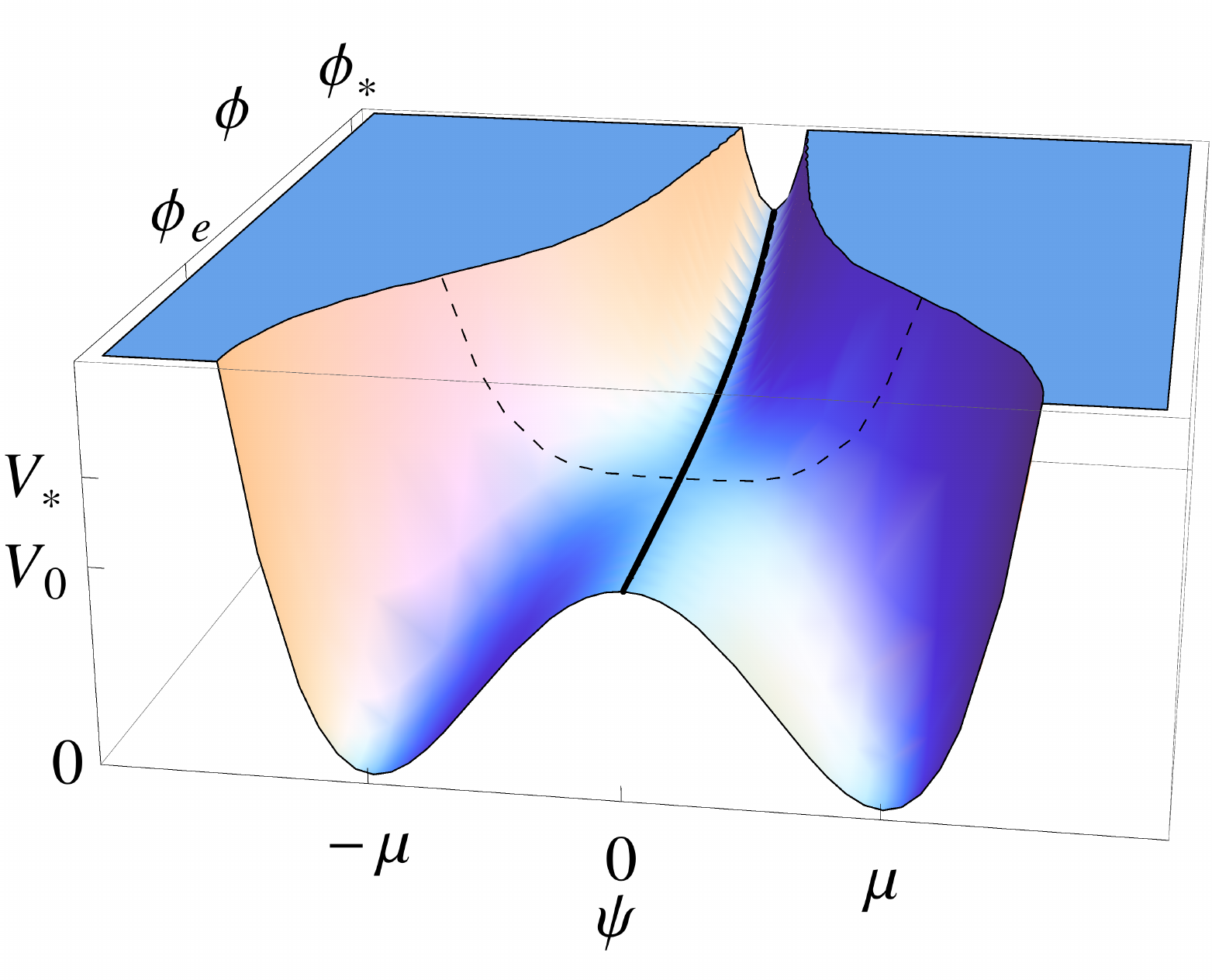}
\end{array}$
  \caption{\textit{Left:} Schematic form of the inflaton potential $V(\phi)$ that can lead to $r \gtrsim 0.1$ with a potential dominated by a false vacuum energy $V_0$. The potential must be steep at $\phi_*$ due to $r = 16 \varepsilon_*$ and then quickly become flat to generate many $e$-folds while staying above the (constant) false vacuum energy $V_0$. Inflation is assumed to end at $\phi_e$ due to a phase transition.\newline  
  \textit{Right:} Illustration of a potential where such a phase transition happens at $\phi_e$ (dashed line) due to an instability in a waterfall field $\psi$ as in hybrid inflation.}
  \label{fig:schematicPotential}
\end{figure}

\section{Why false vacuum energy domination with $r \gtrsim 0.1$ implies small $\Delta \phi$ and a scale-dependent spectral index $\ns(k)$}
\label{sec:analytic}

In this section, we want to discuss why combining $V_0$ domination with a large tensor-to-scalar ratio $r \gtrsim 0.1$ requires relatively small field excursions $\Delta \phi$ and a non-constant ``running'' spectral index $\ns(k)$.

The basis of our discussion is the observation that a large $r$ requires the inflaton potential $V(\phi)$ to be steep at $\phi_*$, while $V_0$ domination requires a potential which is very flat during inflation.
In order to reconcile these conflicting requirements, we must start with a steep potential which very quickly becomes flat during inflation (see fig.~\ref{fig:schematicPotential}). The rapid change in $V'(\phi)$ is then the source of the large running of the spectral index, while the flatness of the potential implies that the inflaton field range takes relatively small values.

\subsection{Flat potential: small $\varepsilon(\phi)$}

We start by formalizing our earlier statement that $V_0$ domination requires a very flat potential. Using eqs.~\eqref{eq:defEpsilon} and \eqref{eq:dNdphi}, we can derive the rate of change of the potential energy $V$ per $e$-fold of slow-roll inflation:
\begin{align}
 \frac{dV}{dN} \, = \, \frac{dV}{d\phi} \frac{d\phi}{dN} \, = \, -2 V \varepsilon. \label{eq:dVdN}
\end{align}
The false vacuum energy $V_0$ is constant during inflation, so only the supposedly negligible inflaton energy density $\tilde{V}$ can contribute to $dV/dN$. Therefore, $V \simeq V_0$ requires that $dV/dN$ must be small, and therefore $\varepsilon$ must be small.

However, the large tensor-to-scalar ratio $r \gtrsim 0.1$ requires a large $\varepsilon_* = r/16$. Such a value of $\varepsilon$ is much too large for $V_0$ domination as one can see from eq.~\eqref{eq:dVdN}. If we had $\varepsilon(\phi) \sim \varepsilon_*$ throughout inflation, the vacuum energy $V_e$ at the end of inflation would be $V_e/V_* \lesssim 50\%$, so $V_0$ (which is constant during inflation) could not have dominated. Therefore, it is necessary that $\varepsilon(\phi)$ quickly decreases during inflation from the large initial $\varepsilon_*$. The faster this decrease happens, the more $V_0$ dominated inflation can be.

\subsection{Scale-dependent spectral index $\ns(k)$}

Such a fast change in $\varepsilon(\phi)$, and therefore $V'(\phi)$, requires that $V''(\phi)$ is large. However, the observed spectral index $\ns \simeq 0.96$ \cite{Planck} requires a small $V''_* \propto \eta_* \sim 10^{-2}$, as one can see from eq.~\eqref{eq:ns}.\footnote{
In principle, $\eta_*$ can be larger if the higher-order slow-roll parameters like $\xi^2_*$ and $\sigma^3_*$ are large, so that they cancel the contribution of a large $\eta_*$ to $\ns$ up to $O(10^{-2})$. However, this would imply that $\ns(k)$ has a strong running according to eqs.~\eqref{eq:alphas} and \eqref{eq:kappas}, so in this case the conclusions of this section remain unchanged.
}
To get a large $V''(\phi)$, it is therefore necessary to have large higher derivatives of $V(\phi)$, and therefore large higher-order slow-roll parameters like $\xi^2$ and $\sigma^3$ according to eqs.~\eqref{eq:defXi} and \eqref{eq:defSigma}.

The higher-order slow-roll parameters are related to the runnings of the spectral index due to eqs.~\eqref{eq:alphas} and \eqref{eq:kappas}, so the requirement to have a strongly varying $\varepsilon$ implies large runnings of the spectral index. Note that it is not strictly necessary that $\as$ and/or $\ks$ are large: one could instead use higher-order runnings related to higher derivatives of $V$. However, one cannot have $V_0$ domination with a truly constant spectral index $\ns(k)$.

The scale dependence of the spectral index is constrained from observations (see appendix \ref{sec:appendix}), and this limits the rate of change of $\eta$ and $\varepsilon$. Also, slow-roll inflation requires that $\eta(\phi) \ll 1$ along the whole slow-roll trajectory, so $\eta$ cannot be increased to arbitrarily high values. For these reasons, some tension between large $r$ and $V_0$ domination remains, and the maximum $V_0/V_*$ for any given $r$ is limited as we will discuss in section \ref{sec:maxVacuumDomination}.

\subsection{Preference for small $\Delta \phi$}

Another consequence of the decreasing $\varepsilon(\phi)$ is that the inflaton field range $\Delta \phi = \lvert \phi_* - \phi_e \rvert$ is relatively small, because $d\phi/dN \simeq \sqrt{2\varepsilon}$ from eq.~\eqref{eq:dNdphi}. Because small $\varepsilon$ is the main condition both for a small inflaton field range and for $V_0$ domination (see eq.~\eqref{eq:dVdN}), larger $V_0/V_*$ is related to smaller field excursions.

Using eq.~\eqref{eq:dNdphi} and $\varepsilon(\phi) \ll \varepsilon_*$ as required by $V_0$ domination, we find that
\begin{align}
 \Delta \phi \, &= \, \int dN \frac{d\phi}{dN}  \, = \, \int dN \, \sqrt{ 2\varepsilon } \, \notag\\
 &\ll \, \int dN \, \sqrt{ 2\varepsilon_* } \, = \, N_e\sqrt{\frac{r}{8}} \, \simeq \, (6\,\mpl)\sqrt{\frac{r}{0.1}}.
\end{align}
This explains why requiring large $r \gtrsim 0.1$ and $V_0$ domination leads to similar conclusions as requiring large $r$ and relatively small field values $\Delta \phi \lesssim \mpl$, for which it has been shown in earlier works \cite{smallField1,smallField2,smallField3} that large running parameters $\as$ and $\ks$ are very helpful.

\section{Estimate for upper bound on $V_0/V_*$}
\label{sec:maxVacuumDomination}

In this section, we want to estimate the maximum possible degree of vacuum energy domination. A useful measure for this is the fraction of energy that is contained in the inflaton field's potential:
\begin{align}
 \frac{\Delta V}{V_0} \, = \, \frac{V_* - V_0}{V_0}.
\end{align}
We can perform a simple estimate of $\Delta V$ that accounts for the two basic limiting factors:
\begin{enumerate}
 \item Initially, $\eta$ is negligibly small, and it takes a while until it increases to significant values, because $V'''$ and higher derivatives are limited by observational constraints on the scale dependence of the primordial spectrum.
 \item Slow-roll inflation requires $\eta(\phi) \ll 1$.
\end{enumerate}
For a rough estimate, we can account for the first limitation by assuming that $\eta = 0$ for the first few $e$-folds,\footnote{
One can easily see that a small initial $\eta_* \sim 10^{-2}$ can be neglected, as going from $\lvert V'_*/V_* \rvert = \sqrt{r/8} \sim 10^{-1}$ to a flat region with $V' \sim 0$ would require roughly $\Delta \phi \sim \lvert V'_*/V''_* \rvert \sim 10$. $\eta_*$ therefore cannot flatten the potential significantly within the small field range $\Delta \phi \ll 6$ required for $V_0$ domination. There is a caveat though: for sufficiently large $\xi^2_*$ and $\sigma^3_*$ (and therefore large $\as$ and $\ks$), the initial value of $\eta_*$ can be larger due to a cancellation in eq.~\eqref{eq:ns}. However, due to the constraints on $\as$ and $\ks$, this effect cannot remove the first stage completely.
} after which we set it to a large $\eta = \eta_2 < 1$.

\subsection{First stage: $\eta \sim 0$}

During the first stage, where $\eta$ is still negligible, the potential energy changes by
\begin{align}
 \frac{ (\Delta V)_1 }{ V_0 }  \, = \, \int \dphi \frac{ V'(\phi) }{ V_0 } \, \simeq \, (\Delta \phi)_1 \sqrt{2 \varepsilon_*} \, = \, (\Delta \phi)_1 \sqrt{ \frac{r}{8} } \, \sim \, \frac{(\Delta \phi)_1}{10},
\end{align}
where $(\Delta \phi)_1$ is the field range over which $\eta$ remains small. Due to the CMB constraints on the running, one typically finds $(\Delta \phi)_1 \gtrsim \mathcal{O}(10^{-1})$. Therefore, the potential energy usually changes by at least a few percent during the first stage.

\subsection{Second stage: $\eta = \eta_2$}

After switching on a large $\eta = \eta_2$ at $\phi=\phi_2$, the potential quickly becomes flat, and inflation ends at $\phi_e$ in the very flat region (see fig.~\ref{fig:schematicPotential}). We can estimate $\phi_e$ as the field value where $\varepsilon(\phi_e) = 0$.\footnote{Because the potential is flat near $\phi_e$, $\Delta V$ is insensitive to the exact value of $\phi_e$, and a rough estimate of $\phi_e$ is sufficient.}

For constant $\eta=\eta_2$ and $V \simeq V_0$, $\varepsilon(\phi)$ is given as
\begin{align}
 \sqrt{2\varepsilon(\phi)} \, = \, -\frac{V'(\phi)}{V} \, \simeq \, -\frac{V'_*}{V_0} - \frac{V''}{V_0} (\phi-\phi_2) \, \simeq \, \sqrt{2\varepsilon_*} - \eta_2 \, (\phi-\phi_2). \label{eq:epsilonPhase2}
\end{align}
The potential becomes flat ($\varepsilon=0$) at
\begin{align}
 \phi_e - \phi_2 \, = \, \frac{\sqrt{2\varepsilon_*}}{\eta_2} \, = \, \frac{1}{\eta_2} \sqrt{\frac{r}{8}}.
\end{align}
Using eq.~\eqref{eq:epsilonPhase2}, we find for $\Delta V$
\begin{align}
 \frac{ (\Delta V)_2 }{ V_0 }  \, &= \, \int\limits_{\phi_2}^{\phi_e} \dphi \frac{ V'(\phi) }{ V_0 } \, 
 \simeq \, \int\limits_{0}^{\phi_e - \phi_2} \dphi \left(  \sqrt{2\varepsilon_*} - \eta_2 \, \phi  \right) \, 
 = \, \left[ \sqrt{\frac{r}{8}} \phi - \frac{\eta_2}{2} \phi^2 \right]^{\phi_e - \phi_2}_{0} \notag\\
 \, &= \,\frac{r}{16 \eta_2} \, \sim \, \frac{10^{-2}}{\eta_2}. \label{eq:deltaVphase2}
\end{align}
As slow-roll inflation requires $\eta_2$ sufficiently smaller than 1, we find that $\Delta V/V_0$ must be at least a few percent.

\subsection{Conclusion for $\Delta V$}

From the above calculations, we see that while slow-rolling down the steeper part of the potential, the vacuum energy typically changes at least by a few percent. We can deduce a hard upper bound using that the contribution from stage 2 is bounded by $\eta < 1$:
\begin{align}
\frac{V_0}{V_*} \, < \, 1 - \frac{r}{16} \, \sim \, 99\%. \label{eq:VeBound}
\end{align}
Achieving the maximum possible $V_0/V_*$ requires
\begin{enumerate}
 \item large higher derivatives of $V$, and therefore a strong scale dependence of the spectral index, to quickly increase $\eta$ and minimise the contribution from stage 1, and
 \item a large maximum slow-roll parameter $\eta$ to minimise the contribution from stage 2.
\end{enumerate}
Both of these require large slow-roll parameters, so for models which push $V_0/V_*$ to its limits, one may want to consider higher-order corrections in the slow-roll expansion. It is not possible to actually saturate the hard bound \eqref{eq:VeBound} without breaking slow-roll due to $\eta \sim 1$ already during the first few $e$-folds after $\phi = \phi_*$.

Note that the arguments in this section are applicable also in the presence of higher-order slow-roll parameters (higher derivatives of $V(\phi)$), particularly the contribution from stage 2, so we expect that our estimate in eq.~\eqref{eq:VeBound} remains valid for any slow-roll potential. It is also straightforward to generalise eq.~\eqref{eq:deltaVphase2} to multi-field models by defining $\varepsilon$ and $\eta$ in terms of derivatives along the inflationary trajectory, so $\Delta V/V_0$ must be at least a few percent even in multi-field models of slow-roll inflation.

\section{Potential reconstruction around $\phi_*$}
\label{quartic}

In order to analyse the effect of a running spectral index $\ns(k)$, we reconstruct the potential $V(\phi)$ around its value at horizon crossing $V_*$ in terms of the spectral index $\ns$, its running $\as$ and its running of the running $\ks$ (for other works on potential reconstruction see \cite{smallField1,smallField2,potentialReconstruction1,potentialReconstruction2}). We then calculate the upper bound on the value of the potential at the end of inflation $V_e\gtrsim V_0$.

Our strategy is based on the fact that cosmological observations can reliably constrain the primordial spectrum only on scales $k\leq k_{\rm NL}$ for which the evolution of perturbations is linear since recombination. On the other end, scales larger than the observable universe are unobservable. Therefore, cosmological observations constrain scales $k_{\rm obs} \in \left[ k_{\rm 0}\,,\, k_{\rm NL}\right]$, where $k_0$ denotes the largest observable scale. The number of $e$-folds of inflation necessary for those scales to exit the horizon is
\be
N_{\rm obs} = \ln \frac{k_{\rm NL}}{k_{\rm 0}}  \sim 8\,.
\en 

After these initial $N_{\rm obs}$ $e$-folds, the inflationary potential is unconstrained. The value of the potential at $N_{\rm obs}$, $V_{\rm obs}$, obeys the hierarchy $V_{\rm obs} > V_e > V_0$. In what follows we derive upper bounds on $V_{\rm obs}$, which we will then translate into upper bounds on $V_0$.

Henceforth we assume that, during the first $N_{\rm obs}$ $e$-folds, higher-order runnings beyond $\ks$ are zero. This implies that the potential around $\phi_*=0$ can be written as
\be
V=V_*\left[1 + \left.\frac{V'}{V}\right|_* \phi + \frac{1}{2}\left.\frac{V''}{V}\right|_* \phi^2+ \frac{1}{6}\left.\frac{V'''}{V}\right|_* \phi^3 + \frac{1}{24}\left.\frac{V''''}{V}\right|_* \phi^4 \right]\,.
\label{potDV}
\en
Expressing the derivatives of the potential in terms of the slow-roll parameters~\eqref{eq:defEpsilon}--\eqref{eq:defSigma}, we find
\be
V=V_*\left[1-\sqrt{2\varepsilon_*}\phi + \frac{1}{2}\eta_*\phi^2 - \frac{1}{6}\frac{\xi^2_*}{\sqrt{2\varepsilon_*}}\phi^3 + \frac{1}{24}\frac{\sigma^3_*}{2\varepsilon_*}\phi^4\right]\,.
\label{potSR}
\en
The slow-roll parameters~\eqref{eq:defEpsilon}--\eqref{eq:defSigma} at horizon crossing are related to the observables $r$, $\ns$, $\as$ and $\ks$ by~\eqref{eq:r}--\eqref{eq:kappas} allowing us to express the potential \eqref{potSR} solely as a function of these observables:
\be
\frac{V}{V_*}=1 - \sqrt{\frac{r}{8}}\phi + \left[\frac{\ns-1}{4}+\frac{3}{32}r + \frac{q_1}{4}\as + \frac{q_1^2-q_2}{4}\ks\right]\phi^2 + \sqrt{\frac{8}{r}}\frac{\as+q_1\ks}{12}\phi^3 +\frac{\ks}{6r}\phi^4\,.
\label{potobs}
\en

The function \eqref{potobs} guarantees that inflation satisfies all constraints on the observables. Indeed, as we argued above, after the first $N_{\rm obs}$ $e$-folds cosmological observations do not constrain the inflationary potential. Therefore we do not specify the potential responsible for the remaining $N_e-N_{\rm obs}$ $e$-folds of inflation but derive an upper bound on $V_e$ given by continuations of \eqref{potobs} that maximise $V_0$ domination. 

Requiring that the inflaton rolls classically throughout inflation and that primordial black holes are not produced too abundantly near the end of inflation implies that $\varepsilon$ has to be larger than \cite{PBH,generalizedSlowroll2}
\be
\varepsilon_{\rm min} \sim 10^{-1} V\,.
\label{diff}
\en
To derive the upper bound on $V_0$, we consider the case in which the transition to small $\varepsilon$ happens immediately after the first $N_{\rm obs}$ $e$-folds, violating slow-roll. Therefore, during the last $N_e-N_{\rm obs}$ $e$-folds, the minimal change in the potential is
\be
\frac{\Delta V}{V_*} \sim (N_e - N_{\rm obs})2\varepsilon_{\rm min}\sim\mathcal{O}(10^{-8})\,,
\en
which is negligible compared to the $\mathcal{O}(10^{-2})$ change during the first $N_{\rm obs}$ $e$-folds. Therefore, although $V_{\rm obs} > V_e > V_{0}$, upper bounds on $V_{\rm obs}$ are effectively equivalent to upper bounds on $V_0$.

\subsection{Numerical analysis}
\label{numan}

In order to obtain an upper bound for the maximum possible false vacuum energy domination, we have evaluated the potential \eqref{potobs} after $N_{\text{obs}}$ $e$-folds, by scanning inflationary trajectories for the ranges
\be
r=0.1\; &{\rm and}&\;r=0.2\,,\nn
\ns&=&0.96\,,\nn
-0.060\leq&\as&\leq0.045\,,\nn
-0.025\leq&\ks&\leq0.090\,.
\label{ranges}
\en

\begin{figure}[ht]
\centering
\subfigure{\includegraphics[height=6.5cm]{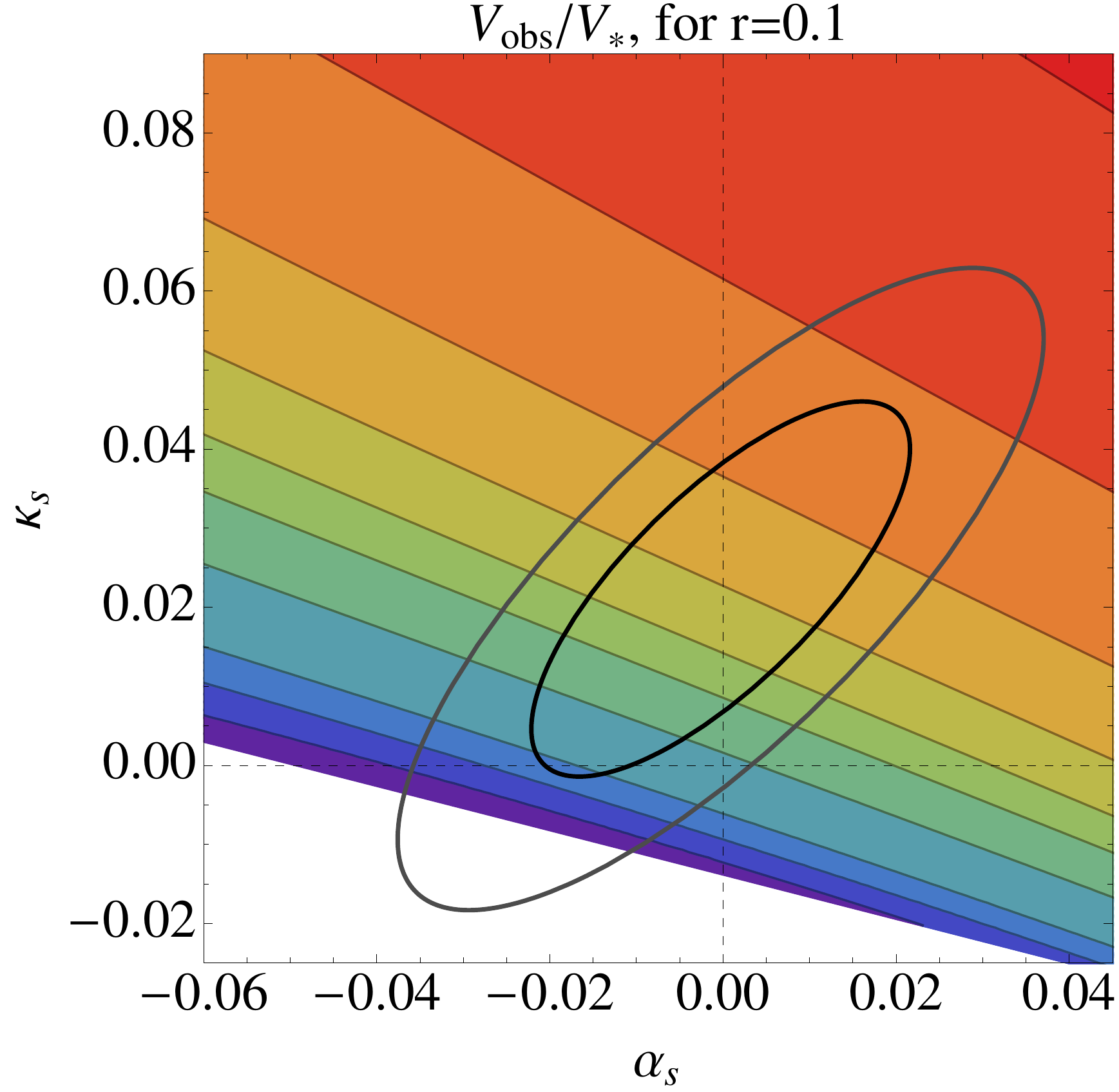}}
\hfill
\subfigure{\includegraphics[height=6.5cm]{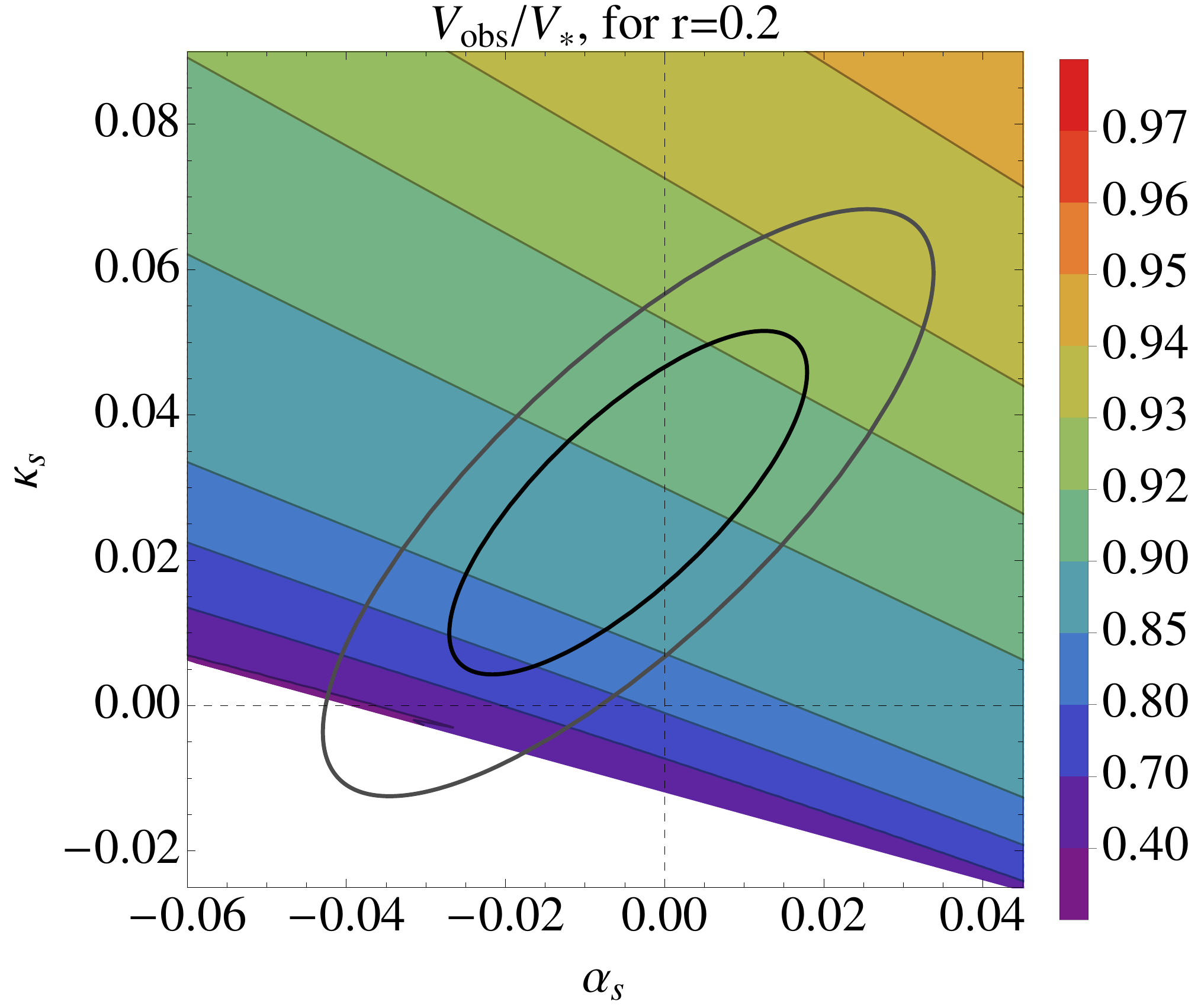}}
\caption{$V_{\text{obs}}/V_{*}> V_0/V_*$ as a function of $\as$ and $\ks$, for $r=0.1$ (left), and $r=0.2$ (right), using $\ns=0.96$. The plots were obtained by evaluating the potential \eqref{potobs} after $N_{\text{obs}}=8$ $e$-folds of inflation. One can see that a larger amount of false vacuum energy domination can be obtained for larger $\ks$ and $\as$. The parameter constraints denoted by the black (68\% CL) and grey (95\% CL) ellipses were obtained using the Planck+BICEP2+BAO data (cf. appendix~\ref{sec:appendix}).}
\label{fig:Vobs}
\end{figure}
\begin{figure}[ht]
\centering
\subfigure{\includegraphics[height=6.5cm]{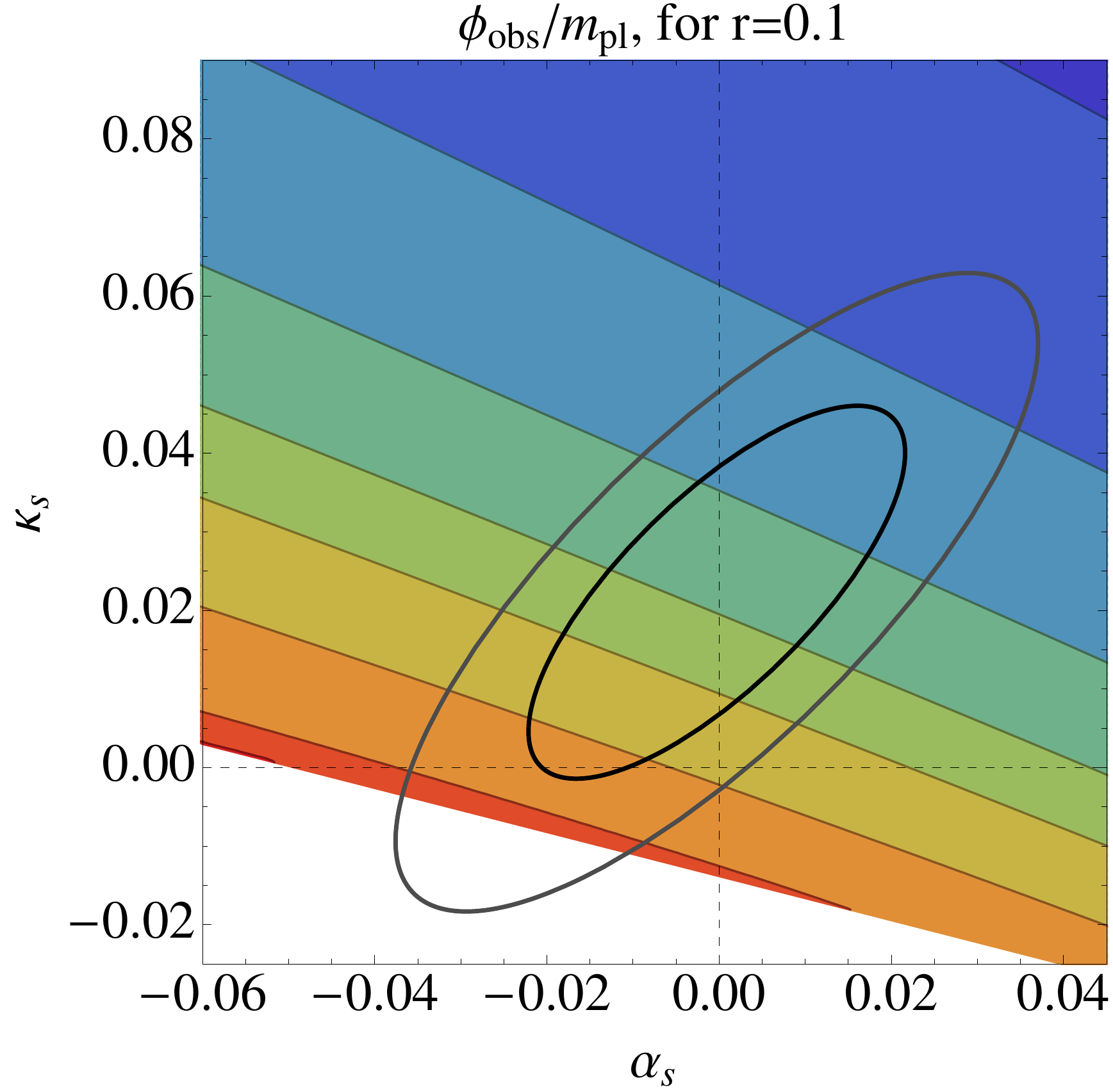}}
\hfill
\subfigure{\includegraphics[height=6.5cm]{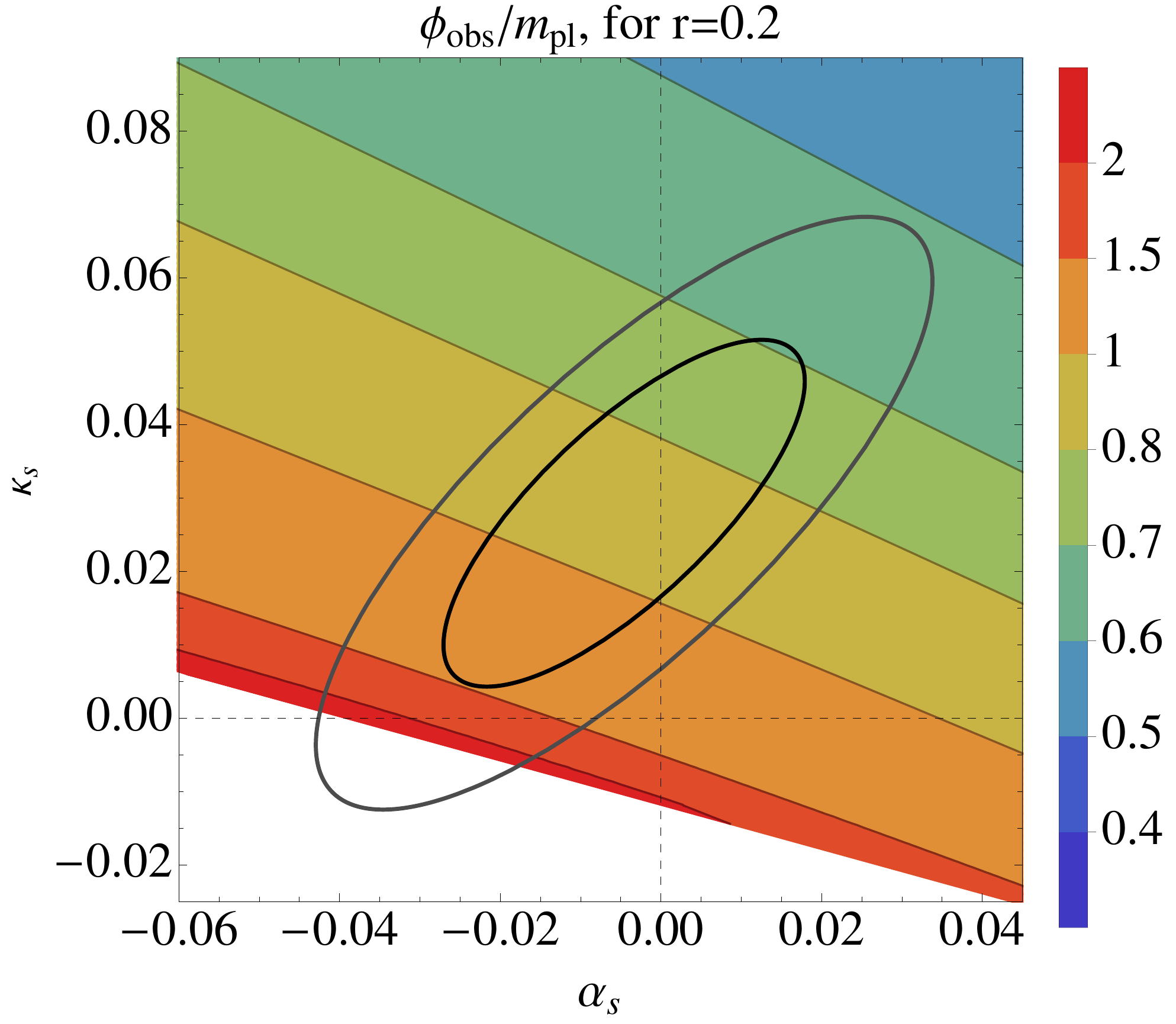}}
\caption{$\phi_{\text{obs}}/\mpl$ as a function of $\as$ and $\ks$, for $r=0.1$ (left), and $r=0.2$ (right), using $\ns=0.96$. Here $\phi_{\text{obs}}$ satisfies the equation $\int_0^{\phi_{\text{obs}}}d\phi V/V'=N_{\text{obs}}$ for the potential \eqref{potobs}. The plots clearly show what we have discussed in section~\ref{sec:analytic}, namely that a larger amount of $V_0$ domination implies that the first $N_{\text{obs}} = 8$ $e$-folds of inflation are obtained within smaller field excursions. The parameter constraints denoted by the black (68\% CL) and grey (95\% CL) ellipses were obtained using the Planck+BICEP2+BAO data (cf. appendix~\ref{sec:appendix}).}
\label{fig:PHIobs}
\end{figure}
\begin{figure}[ht]
\centering
\subfigure{\includegraphics[height=6.5cm]{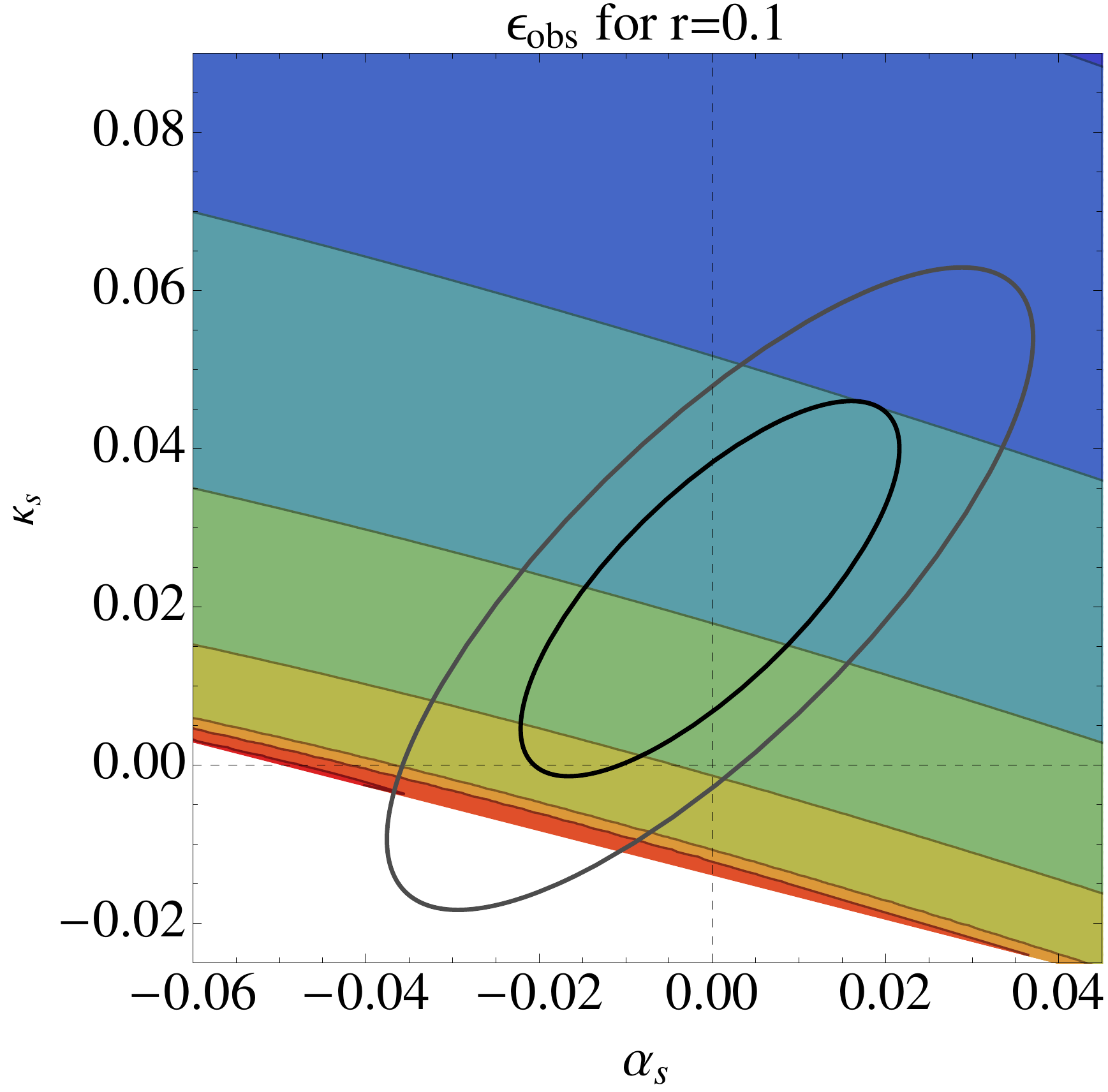}}
\hfill
\subfigure{\includegraphics[height=6.5cm]{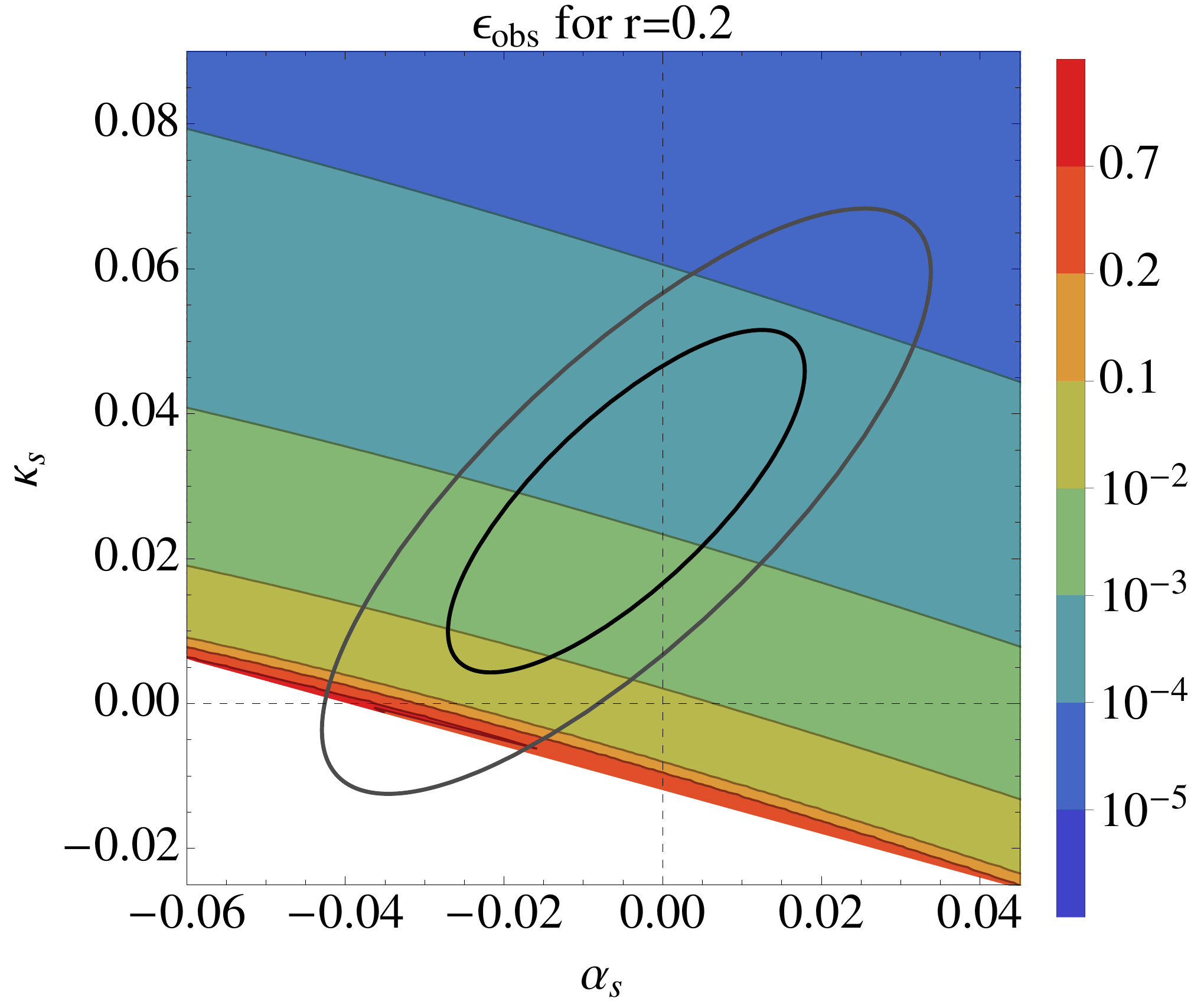}}
\caption{$\varepsilon_{\text{obs}}\equiv \varepsilon(\phi_{\rm obs})$ for r=0.1 (left) and r=0.2 (right) with $\ns=0.96$. For large and positive $\as$ and $\ks$, $\varepsilon_{\rm obs}$ decreases to very small values during the first $N_{\rm obs} = 8$ $e$-folds of inflation whereas for small $\ks$ it remains large. The parameter constraints denoted by the black (68\% CL) and grey (95\% CL) ellipses were obtained using the Planck+BICEP2+BAO data (cf. appendix~\ref{sec:appendix}).}
\label{fig:EPSobs}
\end{figure}

Fig.~\ref{fig:Vobs} shows $V_{\text{obs}}/V_*$ as a function of $\as$ and $\ks$ for $r=0.1$ (left) and $r=0.2$ (right). The corresponding field excursions are illustrated in fig.~\ref{fig:PHIobs} in terms of $\mpl$ and the value of $\varepsilon_{\rm obs}$ is shown in fig.~\ref{fig:EPSobs}. The parameter constraints on $\as$ and $\ks$ are given by the grey (95\% CL) and black (68\% CL) ellipses (cf. appendix~\ref{sec:appendix}).

A first view on fig.~\ref{fig:Vobs} makes clear that a smaller value of $r$ results in an overall increase of the maximum $V_{\text{obs}}/V_*$. Moreover, by comparing fig.~\ref{fig:Vobs} with fig.~\ref{fig:PHIobs} one can see that larger $V_{\text{obs}}/V_*$ corresponds to smaller field excursions, just as expected from our analysis in section~\ref{sec:analytic}.

A closer look at the plots in fig.~\ref{fig:Vobs} reveals the importance of $\ks$ within the scope of a $V_0$ dominated scenario. For no running, i.e. $\as=\ks=0$ we find that $V_{\text{obs}}/V_*\,\lesssim0.89$, for $r=0.1$ and $V_{\text{obs}}/V_*\,\lesssim0.80$ for $r=0.2$ . However, note that this point already lies outside the 95\% CL region for $r=0.2$. Taking $\ks=0$ and $\as<0$ results either in less $V_0$ domination or in inflation ending before the required number of $e$-folds $N_{\text{obs}}$ is achieved. The inflationary trajectories for which $\ks=0$ and $\as>0$ clearly lie outside the 95\% CL.

From the potential \eqref{potobs} one can see that the the spectral index only affects the value of $V''_*$, where larger values of $\ns$ correspond to larger $V''_*$, which in turn leads to an overall increase in $V_{\text{obs}}$. However, we have found that the effect of $\ns$ on $V_{\rm{obs}}$ is in general small, except for the lower-left region of fig.~\ref{fig:Vobs}, where $V_{\rm{obs}}$ is relatively small either way.

\subsection{On the relation between $V_{\rm obs}$, $V_e$ and $V_0$}
\label{obsto0}

In fig.~\ref{fig:Vobs} we show upper bounds on $V_{\rm obs}/V_*$. However, we are ultimately interested in the bounds on $V_0/V_*$. The difference between the two bounds is related to the value of $\varepsilon_{\rm obs}$ after the first $N_{\rm obs}$ $e$-folds of inflation. Fig.~\ref{fig:EPSobs} illustrates that $\varepsilon$ decreases the most for large and  positive $\as$ and $\ks$ while it remains large for small $\ks$. 

Trajectories for which $\varepsilon$ decreases to very small values during the first $N_{\rm obs}$ $e$-folds can remain within the slow-roll approximation for the last $N_e -N_{\rm obs}$ $e$-folds of inflation without losing percentages of false vacuum energy domination. Indeed, in the upper right corner of fig.~\ref{fig:EPSobs} we see that $\varepsilon_{\rm obs} \sim \mathcal{O}(10^{-4})$. Considering the continuation of the potential  that evolves towards flatness as fast as possible within the slow-roll approximation, together with the fact that $V_0\leq V_e$ can be arbitrarily close to $V_e$, implies that the bounds on $V_{\rm obs}/V_*$ are equivalent to bounds on $V_0/V_*$ at the percent level.

On the other hand, trajectories with small $\ks$ for which $\varepsilon_{\rm obs}$ is large lose at least a further few percent of potential energy if slow-roll remains valid throughout inflation, that is $V_0/V_* < 0.99\, V_{\rm obs}/V_*$, as discussed in section~\ref{sec:maxVacuumDomination}. However, if slow-roll is violated and $\varepsilon$ abruptly decreases to very small values $\varepsilon\sim\varepsilon_{\rm min}$ immediately after the first $N_{\rm obs}$ $e$-folds of inflation, the bounds on $V_{\rm obs}/V_*$ are effectively equivalent to bounds on $V_0/V_*$ even for small $\ks$.

\subsection{Effects of higher-order runnings beyond $\ks$}

Our reconstruction of the inflaton potential \eqref{potobs} is based on the assumption that higher-order runnings beyond $\ks$, i.e.\ all $\left[ (d / d \ln k)^{n} \ns(k)\right]_{k=k_*}$ with $n > 2$, are negligible throughout the first $N_{\rm obs}$ $e$-folds. As such higher-order runnings are proportional to the $(n \! + \! 2)$-th derivatives of the inflaton potential $V(\phi)$, this assumption ensures that the inflaton potential can be approximated by a fourth-order polynomial in $(\phi - \phi_*)$. If one allows for higher-order runnings beyond $\ks$, higher-order terms $(\phi-\phi_*)^{n+2}$ with $n>2$ need to be taken into account.

As we discussed in section~\ref{sec:analytic}, to maximise $V_0/V_*$ it is necessary that $V'$ changes rapidly from its large value at horizon crossing to a small value. This change has to be driven by higher-order derivatives of the potential, be it the third and fourth only as in \eqref{potobs} or higher-order derivatives also. Including higher than fourth-order terms and choosing their value at horizon crossing to maximise $V_0$ domination would presumably lead to a larger fraction $V_e/V_*$ than we found in section \ref{quartic}. However, the hard bound \eqref{eq:VeBound} is independent of the form of the potential, being solely due to $r\gtrsim 0.1$ and $\eta < 1$, so the potential has to decrease by at least one percent during inflation in any case.

Note that if the higher-order runnings of the spectral index are not negligible, one cannot rely on cosmological parameter constraints that were derived under the assumption that these runnings are zero -- instead, one should redo the Bayesian parameter estimation for the primordial spectrum including the relevant higher-order runnings to get the correct constraints for that case. In the same way in which the upper bound on $\as$ is relaxed when allowing for $\ks \neq 0$ (see appendix \ref{sec:appendix}), adding large higher-order runnings will likely change the bounds on $\as$ and $\ks$. Higher-order terms also give additional contributions to eqs.~\eqref{eq:ns}--\eqref{eq:kappas}, which change the prefactors of the $\phi^2$, $\phi^3$ and $\phi^4$ terms in eq.~\eqref{potobs} by terms proportional to the additional higher-order runnings.

It may therefore be possible to get large $V_0$ domination even for small $\ks$, but with large higher-order running parameters. In any case, $V_0/V_* > 90\%$ is possible only when there is some sizeable running of the spectral index beyond $\as \neq 0$, e.g.\ $\ks > 0$ or some higher-order running.

\section{Summary and conclusions}
\label{sec:conclusions}
In this paper, we have discussed to which extent and under which conditions false vacuum energy ($V_0$) dominated slow-roll inflation is compatible with $r \gtrsim 0.1$.

We started with general considerations based on the slow-roll dynamics and constraints on the observed spectrum of perturbations. We found that $V_0$ domination requires a strongly scale dependent (``running'') spectral index $\ns(k)$, and that it is related to relatively small inflaton field excursions $\Delta \phi$.

For the maximum amount of $V_0$ domination, we derived a hard bound of $V_0/V_* < 99 \%$ assuming only $\eta < 1$ and $r \gtrsim 0.1$. As slow-roll generally requires $\eta \ll 1$, models of slow-roll inflation will generally remain at least a few percent below this bound.\footnote{We note that some authors (e.g.\ \cite{generalizedSlowroll1,generalizedSlowroll2}) define slow-roll as requiring only $\varepsilon \ll 1$, allowing for $\eta \gtrsim 1$. Our definition of slow-roll also demands that $\eta$ must be small, so that the conventional slow-roll expansion \cite{slowrollExpansion} in $\varepsilon$ and $\eta$ is valid.}

To understand the effect of the running spectral index $\ns(k)$ more quantitatively, we studied a potential reconstruction around the horizon crossing scale $\phi_*$ including the spectral index $\ns$, its running $\as$ and its running of the running $\ks$, which provides a model-independent analysis for the case in which the higher-order runnings beyond $\ks$ are negligible. We calculated upper bounds on $V_0/V_*$ and lower bounds on $\Delta \phi$ as functions of $\as$ and $\ks$ as shown in figs.~\ref{fig:Vobs} and \ref{fig:PHIobs}.

For $\ks = 0$, we found an upper bound $V_0/V_* < 90\%$, whereas for $\ks > 0$, the maximum amount of false vacuum energy domination increases up to $V_0/V_* \simeq 96 \%$. This shows that large $V_0$ domination prefers higher-order runnings beyond $\as$. Adding only $\as$ but no other runnings cannot increase $V_0/V_*$ due to observational constraints on $\as$.

We also derived the joint constraints on $\as$, $\ks$ and $r$ from the combined Planck and BICEP2 likelihoods (see appendix~\ref{sec:appendix}) to correctly constrain $\as$ and $\ks$ in our analysis.\footnote{The constraints on $\ks$ published by the Planck collaboration \cite{Planck} are not directly applicable because they assume $r = 0$.} These constraints can also be used to constrain other models of inflation with $\ks \neq 0$.

We conclude that false vacuum energy domination during slow-roll inflation is possible even with $r \gtrsim 0.1$, but that $V_0/V_* \geq 90\%$ requires higher-order runnings beyond $\as$ whose effects must be included carefully in the study of any such model.

\subsection*{Acknowledgements}
This work was supported by the Swiss National Science Foundation. F.C.\ and D.N.\ thank Benjamin Audren, Julien Lesgourgues and Thomas Tram for the introduction to CLASS and Monte Python during the ``Tools for Cosmology'' workshop in Geneva. We also thank Vinzenz Maurer for helpful discussions.

\appendix
\section*{Appendix}
\section{Constraints on $\as$ and $\ks$ from Planck and BICEP2}
\label{sec:appendix}

\renewcommand{\arraystretch}{1.6}% array stretch factor
\begin{table}[tbph]
\centering
  \begin{tabular}{ | c | c | c | c | }
\hline
  & $r$ fitted & $r=0.1$ fixed &  $r=0.2$ fixed \\
\hline
  $n_{\rm s}$ & $0.961_{-0.006}^{+0.006}$ & $0.961_{-0.006}^{+0.006}$ & $0.961_{-0.006}^{+0.006}$ \\
  $\alpha_{\rm s}$ & $-0.005_{-0.016}^{+0.016}$ & $0.000_{-0.016}^{+0.016}$ & $-0.004_{-0.016}^{+0.016}$\\
  $\kappa_{\rm s}$ & $0.029_{-0.018}^{+0.017}$ & $0.023_{-0.017}^{+0.017}$ & $0.029_{-0.017}^{+0.017}$ \\
  $r$ & $0.208_{-0.048}^{+0.040}$ & $0.1$ & $0.2$\\
  $A_{\rm s}$ & $\left(  2.28_{-0.07}^{+0.06}  \right)\times 10^{-9}$ & $\left(  2.27_{-0.7}^{+0.6}  \right)\times 10^{-9}$& $\left(  2.28_{-0.07}^{+0.06}  \right)\times 10^{-9}$ \\
  \hline
  $\omega_{\rm b}$ & $0.0223_{-0.0003}^{+0.0003}$ & $0.0223_{-0.0003}^{+0.0003}$& $0.0223_{-0.0003}^{+0.0003}$\\
  $\omega_{\rm cdm}$ & $0.1170_{-0.0014}^{+0.0014}$ & $0.1172_{-0.0014}^{+0.0014}$& $0.1170_{-0.0014}^{+0.0014}$\\
  $H_0$ & $\left( 68.71_{-0.69}^{+0.66} \right)\operatorname{\frac{km}{s \, Mpc}}$ & $\left( 68.53_{-0.67}^{+0.65} \right)\operatorname{\frac{km}{s \, Mpc}}$& $\left( 68.72_{-0.68}^{+0.66} \right)\operatorname{\frac{km}{s \, Mpc}}$\\
  $\tau_{\rm reio}$ & $0.110_{-0.015}^{+0.014}$ & $0.109_{-0.015}^{+0.014}$ & $0.110_{-0.016}^{+0.014}$\\
  \hline
  \end{tabular}
\caption{Cosmological parameter constraints for Planck + BICEP2 + BAO at 68\% CL with the primordial spectrum expanded around the pivot scale $k_* = 0.05$ Mpc$^{-1}$. Note that $\kappa_{\rm s}$ has a strong correlation with $\alpha_{\rm s}$ which should be taken into account when applying these constraints, see fig.~\ref{fig:MontePythonTriangle}.}
\label{tab:MontePython}
\end{table}

\begin{figure}[tbph]
\includegraphics[width=\textwidth]{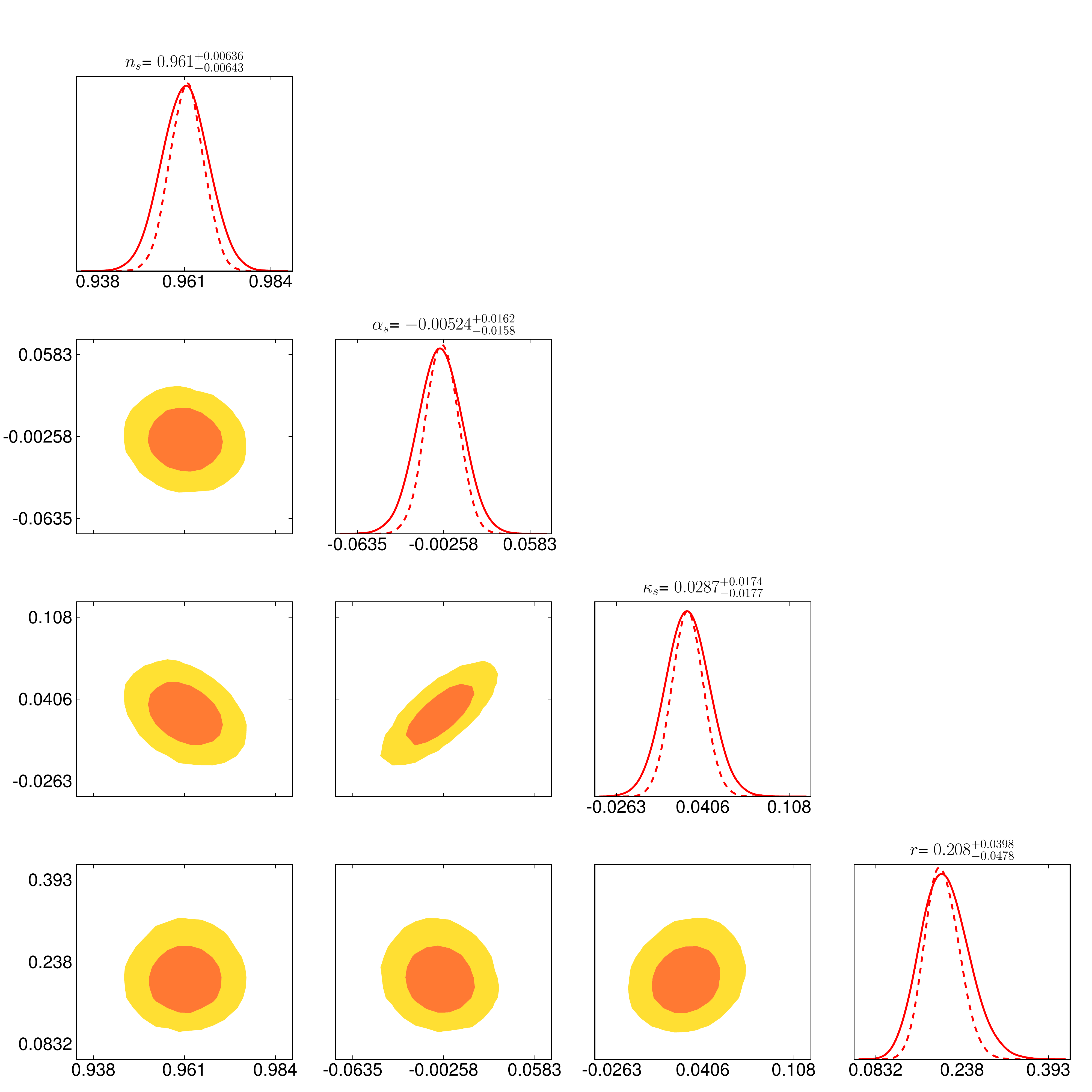}
\caption{Constraints on the primordial spectrum expanded around the pivot scale $k_* = 0.05$ Mpc$^{-1}$ for Planck + BICEP2 + BAO, assuming a $\Lambda$CDM model extended by $r$, $\as$ and $\ks$ (see eqs.~\eqref{eq:ansatzPs} and \eqref{eq:ansatzPt}).}
\label{fig:MontePythonTriangle}
\end{figure}

As the constraints on $\as$ and $\ks$ provided by the Planck collaboration \cite{Planck} assume a vanishing tensor-to-scalar ratio $r=0$, they are not applicable for large $r \gtrsim 0.1$. Since the BICEP2 data has been released, several independent Bayesian parameter estimates for $\as$ and $r$ using the joint Planck and BICEP2 likelihoods have appeared \cite{BICEP2fit1,BICEP2fit2,BICEP2fit3} which find some evidence for a negative running $\as < 0$.\footnote{For parameter estimates using different parametrizations of the primordial spectrum, see \cite{BICEP2fitAlt1,BICEP2fitAlt2}.} However, all of these analyses assume that $\ks = 0$, and to our knowledge no parameter estimation including both $\ks$ and $r \neq 0$ has been published yet.

To derive the joint constraints on $\Lambda$CDM + $r$ + $\as$ + $\ks$ from current observations, we have used the cosmological parameter estimation code Monte Python \cite{MontePython} together with the Boltzmann code CLASS \cite{CLASS1,CLASS2}, assuming primordial scalar and tensor spectra of the form
\begin{subequations}
\begin{align}
 \mathcal{P}_{\rm{s}} \, &= \, A_{\rm s} \left( \frac{k}{k_*} \right)^{ \ns-1 \,+\, \frac{\as}{2}\ln\left( k/k_* \right) \,+\, \frac{\ks}{6}\ln^2\left( k/k_* \right) },  \label{eq:ansatzPs}\\
 \mathcal{P}_{\rm t} \, &= \, r \, A_{\rm s} \left( \frac{k}{k_*} \right)^{-r/8}, \label{eq:ansatzPt}
\end{align}
\end{subequations}
with the arbitrary pivot scale $k_* = 0.05$ Mpc$^{-1}$. Note that we have enforced the slow-roll consistency condition $n_{\rm t} = -r/8$ because we are interested in parameter constraints for slow-roll inflation only. The parameter constraints are derived using the Planck likelihood data from the Planck Legacy Archive and the BICEP2 \cite{bicep2} and BOSS BAO \cite{bao} likelihood data as included in Monte Python 2.0.4.

The resulting 68\% CL parameter constraints for $r$ as a free parameter and for fixed $r=0.1$ and $r=0.2$ are given in table \ref{tab:MontePython}, and a triangle plot of the 2-dimensional constraints for the primordial spectrum for fitted $r$ is shown in fig.~\ref{fig:MontePythonTriangle}. We find a strong correlation between $\as$ and $\ks$, similar to the Planck constraints for the $r=0$ case \cite{Planck}, but with a slight shift towards larger $\ks$ and smaller $\as$. For $\ks = 0$, we find $\as < 0$ in good agreement with \cite{BICEP2fit1,BICEP2fit2,BICEP2fit3}\footnote{Our results cannot be compared with \cite{BICEP2fitAlt1} because they treat $n_{\rm t}$ as a free model parameter, whereas we restrict our analysis to slow-roll spectra with $n_{\rm t} = -r/8$.}. However, for $\ks > 0$, we find that $\as > 0$ is preferred due to the strong correlation between $\as$ and $\ks$.

\end{document}